%% file: main.tex
\documentclass[prd,showpacs,letterpaper,twocolumn,nofootinbib,longbibliography,floatfix,superscriptaddress]{revtex4-1}
\input{Packages}

\input{Commands}
\begin{document}

% \preprint{APS/123-QED}
\begin{CJK*}{UTF8}{gbsn}
\title{Comparison of Geometrical Layouts for Next-Generation\\ Large-volume Cherenkov Neutrino Telescopes}
%\title{Optimization of Geometrical Layout for Next-Generation Water/Ice Cherenkov Neutrino Telescopes Using Graph Neural Networks}% Force line breaks with \\
%\thanks{A footnote to the article title}%

\author{Tong Zhu~(朱彤)\orcidA{}}
 \email{zhutong@berkeley.edu}%Lines break automatically or can be forced with \\
\affiliation{Department of Physics, University of California, Berkeley, CA 94720, USA}
 
\author{Miaochen Jin~(靳淼辰)\orcidB{}}
\email{miaochenjin@g.harvard.edu}
\affiliation{Department of Physics \& Laboratory for Particle Physics and Cosmology, Harvard University, Cambridge, MA 02138, USA}

\author{Carlos A.~Arg{\"u}elles\orcidC{}}
\email{carguelles@fas.harvard.edu}
\affiliation{Department of Physics \& Laboratory for Particle Physics and Cosmology, Harvard University, Cambridge, MA 02138, USA}

\date{\today}% It is always \today, today,
             %  but any date may be explicitly specified

\begin{abstract}
Water-(Ice-) Cherenkov neutrino telescopes have played a pivotal role in the search and discovery of high-energy astrophysical neutrinos.
Experimental collaborations are developing and constructing next-generation neutrino telescopes with improved optical modules (OMs) and larger geometrical volumes to increase their efficiency in the multi-TeV energy range and extend their reach to EeV energies.
Although most existing telescopes share similar OM layouts, more layout options should be explored for next-generation detectors to maximize discovery capability.
In this work, we study a set of layouts at different geometrical volumes and evaluate the signal event selection efficiency and reconstruction fidelity under both an only trigger-level linear regression algorithm and an offline Graph Neural Network (GNN) reconstruction. 
Our methodology and findings serve as first steps toward an optimized, global network of neutrino telescopes.
\end{abstract}
\maketitle
\end{CJK*}
% \tableofcontents

\section{Introduction}\label{sec:intro}

More than ten years ago, the IceCube Neutrino Observatory observed high-energy astrophysical neutrinos~\cite{IceCube:2014stg}, opening a new window into the universe.
Since then, a plethora of neutrino telescopes have been proposed~\cite{Ackermann:2022rqc} to follow up IceCube measurements and to expand its capabilities.
Neutrino telescopes~\cite{Klein:2022lrf} are gigaton-scale neutrino detectors that use naturally occurring media such as glaciers~\cite{ANITA:2016vrp,IceCube-Gen2:2020qha}, lakes~\cite{Dvornicky:2024ujq}, sea~\cite{KM3Net:2016zxf,Ye:2022vbk,P-ONE:2020ljt}, interstellar dust~\cite{POEMMA:2020ykm}, or mountains~\cite{Adams:2017fjh,Thompson:2023pnl,Otte:2023osf} as neutrino targets.

The majority of existing and proposed optical-Cherenkov neutrino telescopes adhere to a design with a rich history~\cite{Spiering:2012xe}, tracing back to the DUMAND project~\cite{Roberts:1992re}.
They consist of several thousands of optical modules (OMs) deployed in liquid or solid water, which detect Cherenkov light emitted by relativistic charged particles produced in neutrino interactions.
The photon arrival times at the different detectors are then used to reject background, identify neutrino events, and infer their direction, energy, and flavor. 
This is the type of detectors, which we will call ``neutrino telescopes,'' will be the primary focus of this article, leaving Earth-skimming neutrino experiment using particle~\cite{Thompson:2023pnl}, optical~\cite{Otte:2023osf,Adams:2017fjh,POEMMA:2020ykm}, and radio~\cite{GRAND:2018iaj} detection beyond our discussion as their layouts, backgrounds, and detection methods are very different.

Though a handful of high-energy astrophysical neutrino sources has been identified~\cite{IceCube:2018cha,IceCube:2022der} and more recently the galactic plane has been detected in neutrinos~\cite{IceCube:2023ame}, the field of high-energy neutrino astrophysics requires detectors that are an order of magnitude larger to make significant progress~\cite{IceCube-Gen2:2020qha,Ye:2022vbk}, and answer pressing questions in physics and astrophysics~\cite{Arguelles:2024xkx}.
Since next-generation neutrino telescopes are under development, discussions on optimal detector geometry are timely and necessary to increase the neutrino event quality and expected sample size.
Currently, few results about the optimal configuration of neutrino detectors can be found in the literature.
A study conducted by the IceCube-Gen2 collaboration~\cite{IceCube-Gen2:2021tmd} focuses on maximizing the point source sensitivity of IceCube-Gen2, a proposed extension of IceCube, by optimizing the spacing between the newly deployed strings and associated optical sensors.
The results of this study provide insights into the design and deployment strategy of IceCube-Gen2, with the ultimate goal of achieving a five times better point source sensitivity compared to the current IceCube detector.
Additionally, recent efforts by P-ONE collaborators towards machine-learning approaches to neutrino telescope optimization are underway~\cite{Haack:2023uwd}.

\begin{figure*}[t]
  \centering
  \includegraphics[width=1\textwidth]{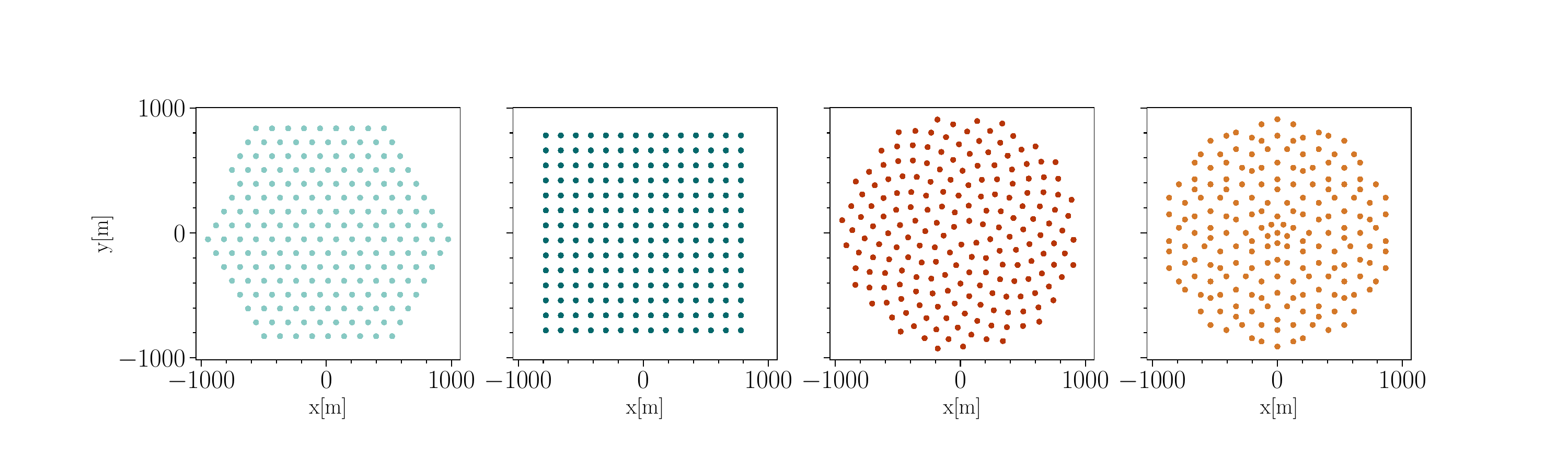}
  \caption{\textbf{{Detector layouts considered in this work.}} 
  From left to right: hexagonal with $\SI{128}\m$ interstring separation (first left), orthogonal with $\SI{120}\m$ interstring spacing (second left), sun flower with $\SI{117}\m$ interstring separation (third left), and Penrose with $\SI{112}\m$ spacing (fourth left).
   For the Penrose geometry, the red dot refer to the five added strings to ensure string number uniformity while keeping quintic rotational symmetry.
   }
  \label{fig:geos}
\end{figure*}

What is the optimal neutrino telescope design is, to a great extent, a matter of scientific taste.
The typical benchmark is the sensitivity to neutrino point sources, whose discovery is the primary goal of neutrino telescopes since it allows detailed comparison with source models often using multi-messenger data~\cite{Murase:2018iyl,Petropoulou:2019zqp,Halzen:2019qkf,Ajello:2023hkh} and opens opportunities to study very long baseline neutrino oscillations~\cite{Keranen:2003xd,Beacom:2003eu,Esmaili:2009fk,Shoemaker:2015qul,Carloni:2022cqz}.
However, one could, e.g., argue that the measurement of the different neutrino flavors is one of the most important goals of neutrino astrophysics particle physics, since it explores neutrino oscillations in unprecedented scales~\cite{Arguelles:2015dca}.
This has, in fact, been demonstrated by the current IceCube array, which has performed some of the most sensitive tests of Lorentz symmetry~\cite{IceCube:2017qyp}.
Such a design will require balancing the capacity to reliably reconstruct various neutrino flavors, which requires closer detector spacing and excellent photon time resolution, putting this in tension with achieving greater effective volumes due to increased instrumentation cost.
One could also argue that the most important goal is to measure the diffuse astrophysical neutrino spectra~\cite{IceCube:2020wum}, since spectral features and comparison to diffuse gamma-ray emission can yield information on the source population~\cite{Murase:2015xka,Bechtol:2015uqb} or allow for novel tests of new physics~\cite{Araki:2014ona,Ng:2014pca,Carpio:2021jhu,IceCube:2023ies}.
Each different scientific goal will yield a different detector optimization. 
This observation is well-aligned with the shift of paradigm from the now, anachronistic, \textit{one detector to rule them all} to a \textit{symbiotic ecosystem of neutrino telescopes}~\cite{Soto:2021vdc}.

In this context, we aim to explore the efficiencies and trigger-level resolutions for different next-generation neutrino telescope geometries when using ice or water as a medium.
Our rationale for studying the detector response at the trigger level follows from the discussion above: higher levels of processing are associated with more specific scientific goals.
Improvements at the trigger level positively improve all potential science cases.
We will additionally focus on the signal efficiency, i.e., the capacity to detect neutrinos, and leave for future work the incorporation of background rejection as an optimization criterion.
This is due to computational limitations associated with the production of background Monte Carlo. 

The rest of this article is organized as follows.
In~\Cref{sec:datasim}, we will first discuss the simulation employed in this work.
We describe the set of sixteen detector geometries, the simulation tools, event distributions, event selection criteria, and effective areas of each setup.
In~\Cref{sec:recomethod}, we discuss the two reconstruction methods we use as benchmarks.
The first method is a simple regression applied at the trigger level, and the second one is a GNN-based machine learning algorithm applied at events of higher quality.
These two methodologies mimic what is currently done on neutrino telescopes, where simpler reconstructions are run on-site and more expensive ones are applied off-line to the data at various levels of data selection.
\Cref{sec:results} presents results on the effective areas and the resolutions of the reconstructions on each candidate detector in this work.
Finally, in~\Cref{sec:summary}, we conclude.

\section{Data Simulation\label{sec:datasim}}

\subsection{Detector Geometry Candidates}

For this work, we consider four different geometrical layouts for OM-loaded strings, as shown in~\Cref{fig:geos}.
Each detector layout is studied at four detector geometrical volumes,  yielding a total of 16 potential detector designs.
To fairly compare the efficiencies of not only different geometries but also various instrumented volumes, we fix the number of OMs as well as the number of strings across all 16 layouts.
This implies larger string separation for larger geometrical volumes and slightly different mean string separations for different geometric layouts when they share approximately the same detector volume.
All candidate configurations consist of 196 strings, each housing 80 OMs with a vertical spacing of $\SI{15.6}\m$. While this depth is comparable with many current-generation neutrino telescopes, we design our candidates to have much larger horizontal cross-sectional geometric areas since next-generation neutrino detectors will likely look to expand horizontally while being limited vertically by the geography of the detector locations. The effect of this limited expansion only sideways will be discussed later in~\Cref{sec:results}.
For this work, we consider a telescope in Antarctic ice, where the mean absorption length for Cherenkov photons is $\sim\SI{250}\m$ for an analogous analysis in water would require scaling down the inter-OM and inter-string distances to account for the typically shorter water absorption length, $\sim\SI{60}\m$.
We expect our conclusions regarding the optimal geometry to be unaffected by this rescaling and the angular resolutions to be improved in the water due to the larger scattering length compared to the absorption length.

In~\Cref{fig:geos}, a top-view of the configurations is shown where each dot represents a detector string.
The hexagonal grid (a) is a shape similar to the IceCube main array, but for this work, we simulate a detector with a perfectly equilateral hexagonal layout, where the uneven outer layer is due to the restriction we set on a total number of strings employed.
The orthogonal grid (b) represents the simplest 2D array configuration.
The sunflower shape (c) is inspired by the baseline design of IceCube-Gen2, which is expected to mitigate corridor events --- background muons that enter the detector through the string spacing; the analytical form in polar coordinates can be found in Ref.~\cite{IC2opt}. 
Lastly, we study the Penrose tilling shape (d) based on the proposed design of TRIDENT, with an additional five strings placed at the center to maintain the exact string count while preserving quintic rotation symmetry.

\begin{figure}[h]
  \centering
  \includegraphics[width=1\textwidth]{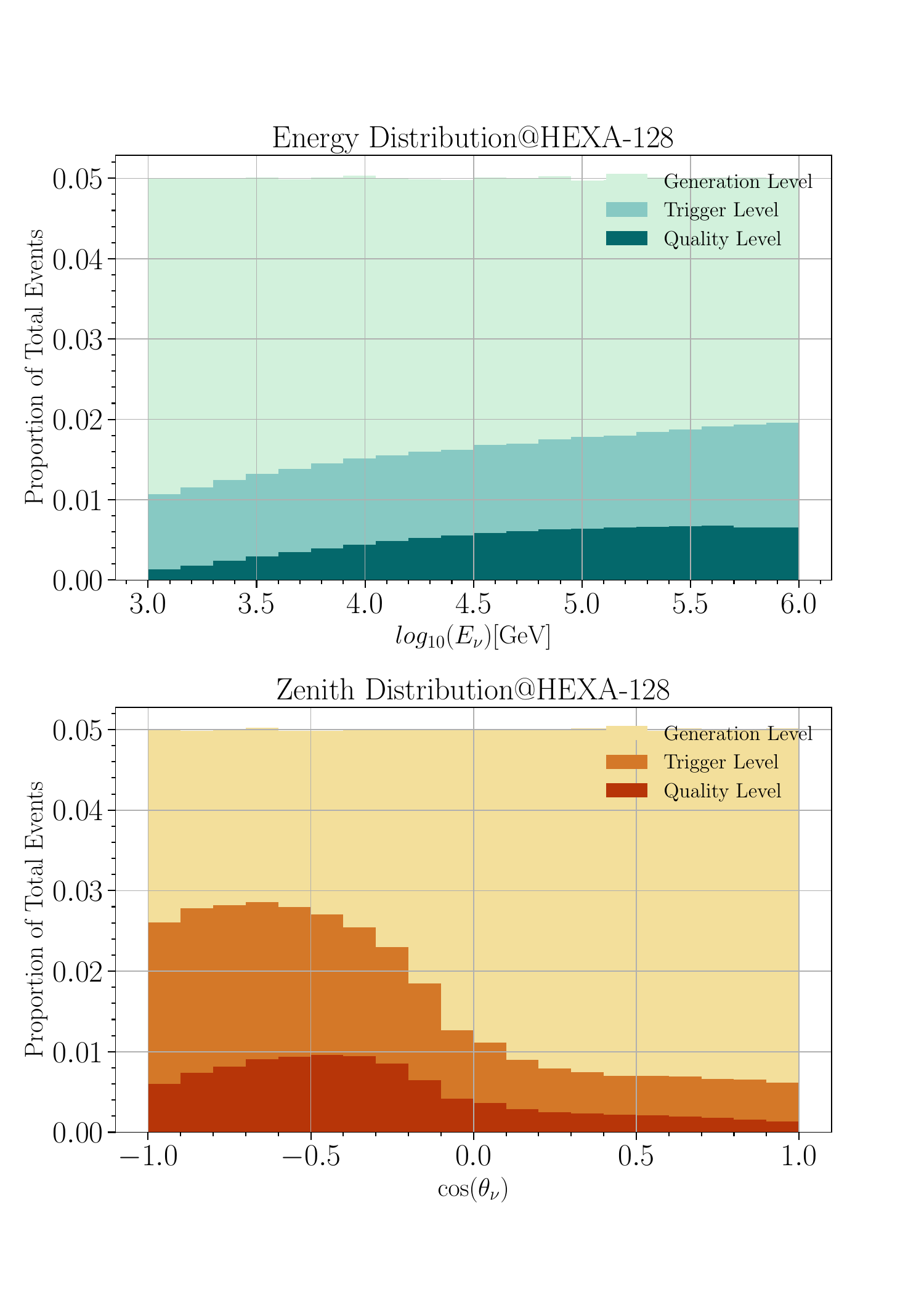}
  \caption{\textbf{Generation, trigger, and quality level event selection for the 128-meter-spacing hexagonal (\texttt{hexa128}) geometry.} Event distributions for other geometry candidates follow similar patterns.}
  \label{fig:dist}
\end{figure}

\begin{table}[h]
    \centering
    \caption{
    $\mathcal{D}_2$ in meters for the 16 different geometries considered in this article. 
    Each line corresponds to a different string configuration, while each column gives the $\mathcal{D}_2$ value.
    Additionally, the top row shows the common geometrical area shared between the different configurations.
    }
    \begin{tabular}{|l|l|l|l|l|}
        \hline
         \diagbox{Configuration}{Area (km$^2$) } & 2.4    & 5.5    & 9.7    & 15.2   \\ \hline
        Hexagonal (\texttt{hexa})                 & 128 & 192 & 256 & 320 \\ \hline
        Orthogonal (\texttt{ortho})                & 120 & 180 & 240 & 300 \\ \hline
        Sunflower (\texttt{sun})                & 117 & 176 & 234 & 293 \\ \hline
        Penrose (\texttt{pen})                  & 112 & 168  & 224 & 280 \\ \hline
    \end{tabular}
    \label{tbl:sixteen-geometries}
\end{table}

To investigate which configuration is more efficient, we demand that the four configurations share the same geometrical volume.
We achieve this by setting the same geometric area in the $x$-$y$ plane.
Additionally, we explore these four configurations in four different scales, corresponding to different geometrical volumes.
In each of these geometries, the lattice points in the $x$-$y$ plane of different configurations are proportionally scaled up.
Since the string spacing is not the same between every pair of neighboring strings, we introduce the following quantity to quantify the horizontal string spacing in these geometries,
\begin{equation}
    \mathcal{D}_2 := \frac 1 {N_{\rm str}}\ \sum^{N_{\rm str}}_{i = 1}\   \frac12 \sum_{j\in N(i)} r(i,j),
  \label{eq2}
\end{equation}
where $N_{\rm str}$ is the total number of strings, which we fix to 196, $N(i)$ are the set of indices of the two-nearest neighbors of the $i_{th}$ string, and $r(i,j)$ is the euclidean distance between the $i_{th}$ and $j_{th}$ strings measured in the $x$-$y$ plane. This metric reflects the scale of the inter-string distance. 
For example, $\mathcal{D}_2 \sim \SI{120}\m$ indicates an array with a sparsity level comparable to the IceCube main array.
Next-generation neutrino telescopes that aim to explore larger energies than existing detectors but employ a similar amount of strings will require a sparser distribution, i.e., a larger $\mathcal{D}_2$ value.
This is the case of IceCube-Gen2, whose benchmark design has a $\SI{250}\m$ inter-string distance in ice.
To facilitate comparison with IceCube and IceCube-Gen2, we assume all detectors to be placed in Antarctic ice.
Our four comparand scales will span the spacing from IceCube to above its planned successor, i.e., the smallest $\mathcal{D}_2$ is set to $\sim \SI{120}\m$ and the largest $\mathcal{D}_2$ to $\sim \SI{300}\m$.
In what follows, we will denote each of the considered geometries by their respective configuration shorthand (\texttt{hexa}, \texttt{ortho}, \texttt{sun}, and \texttt{pen}) and corresponding $\mathcal{D}_2$ values, which can be found on~\cref{tbl:sixteen-geometries}.

\subsection{Event Simulation and Selection}

For this work, we use the open-source neutrino telescope simulation \texttt{Prometheus}~\cite{jeff}.
We simulate muon-neutrino events between $10^3$ to $10^6$~GeV according to an unbroken power law with a spectral index of $-1$.
Events are injected using the ranged injection mode with a uniform radius in column density with a cutoff at the top of the atmosphere, while the angular distribution is uniform in cosine zenith; this results in more down-going than up-going events, as shown in~\Cref{fig:dist}.
We apply two levels of cut to the events, which we call \textit{trigger} and \textit{quality}.
At the trigger level, we apply a coincidence trigger selection criteria similar to those described in Ref.~\cite{trigger,felix,jin2023watts}.
Specifically, we require the observation of twelve occurrences within five milliseconds of coincidence detection in neighboring or next-to-neighboring OMs, where each pair of coincidental photon deposition must be observed within a five-millisecond time window.
This ``trigger level'' selection mimics the one used in IceCube, which selects events that deposit enough amount of light such that they can be distinguished from the background and thereby can be evaluated using fast online reconstruction methods as a first filter. 
Additionally, we also implement a ``quality level'' that further discards events that hit too few OMs to leave a clear morphological signature.
The main objective of this selection is to remove the so-called ``corner-clippers,'' where the muon track only passes through a small fraction of the detector, and to remove events that are either too spherical or too elongated to be identified as muon tracks in a morphological classifier.
Specifically, the quality level selection contains three separate criteria:
\begin{eqnarray}
N_{\rm OM} > 45, \nonumber \\
r_{\rm C} < 0.7\times R_{\rm det}, \\
z_{\rm C} < 0.8\times Z_{\rm det}, \nonumber\\
2<R_{\rm ell}<7 \nonumber
\end{eqnarray}
where $N_{\rm OM}$ is the number of OMs that detected light, $r_{\rm C}$ is the horizontal distance between the average photon position and the center of our detector, $R_{\rm det}$ is the maximum horizontal distance between strings and the center of our detector, $z_{\rm C}$ is the vertical distance between the average photon position and the center of our detector, $Z_{\rm det}$ is the half-height of our detector and $R_{\rm ell}$ is the ratio of the long-axis length to the short-axis length of the ellipsoid fit to the photon hits.

\begin{figure}[ht]
  \centering
  \includegraphics[width=1\textwidth]{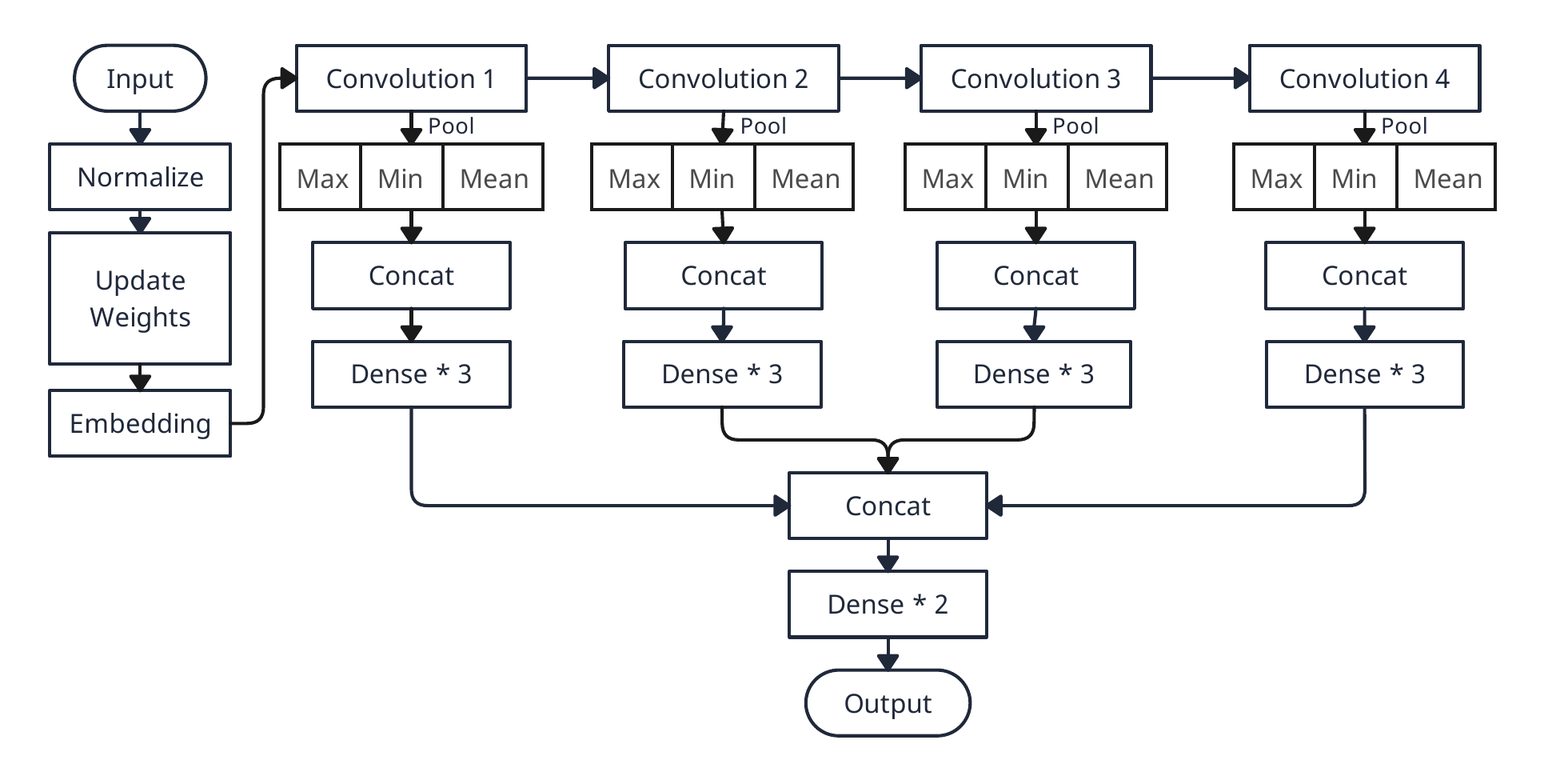}
  \caption{\textbf{Graph Neural Network Architecture Design.}}
  %\note{}
  \label{fig:archi}
\end{figure}

\begin{figure*}[t]
  \centering
  \includegraphics[width=1\textwidth]{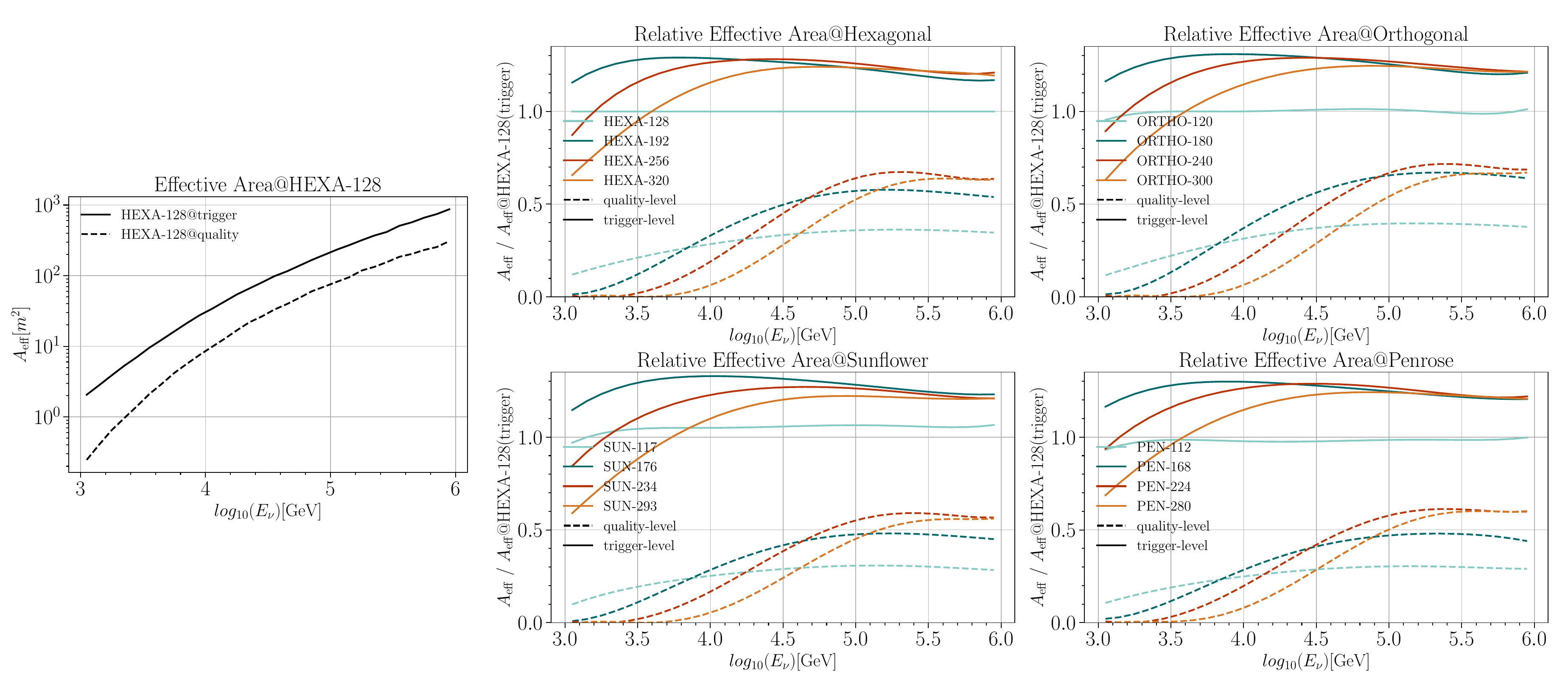}
     \caption{\textbf{Effective areas of detector candidates.} \textbf{Left:} The trigger-level  and quality-level effective area of \texttt{hexa128}. \textbf{Right:} Relative effective areas are depicted with solid lines for trigger-level and dashed lines for quality-level. We take the trigger-level effective area of \texttt{hexa128} as a benchmark to better illustrate the differences of various geometries at different geometric volumes.}

  \label{fig:Aeff}
\end{figure*}

\section{Reconstruction Method}\label{sec:recomethod}

We study the potential of angular reconstruction resolution at both selection stages.
At trigger level, we apply a simple regression fit, which we will denote as ``\texttt{LineFit},'' to the OM photon hits as
\begin{equation}\label{eq:linefit}
    \min_{t_0, \vec{x}_0, \vec{v}} \sum_{i = 1}^N ||\vec{v}(t_i - t_0) + \vec{x}_0 - \vec{x}_i||.
\end{equation}
This reconstruction mimics the first-guess reconstruction used in current detectors and applies to scenarios of real-time online data processing where only computationally light regressions are used.

At the quality selection level, we use a Graph Neural Network (GNN) to perform convolution on the photon hits, similar to Convolutional Neural Network (CNN), which has recently been used by IceCube~\cite{IC2opt}. In our method, each event is converted to a graph with directional edges.
By replacing the conventional convolution with graph convolution, we are able to encode the irregular OM hit spatial distribution.
Furthermore, we categorize node links (edges) into different groups to promote efficiency in information passing between graph vertices.
The details of data preprocessing and the graph schema design will be discussed in~\Cref{subsec:schema}.
Here, we present and discuss the neural network architecture design shown in~\Cref{fig:archi}.

\begin{figure*}[t]
  \centering
  \includegraphics[width=1\textwidth]{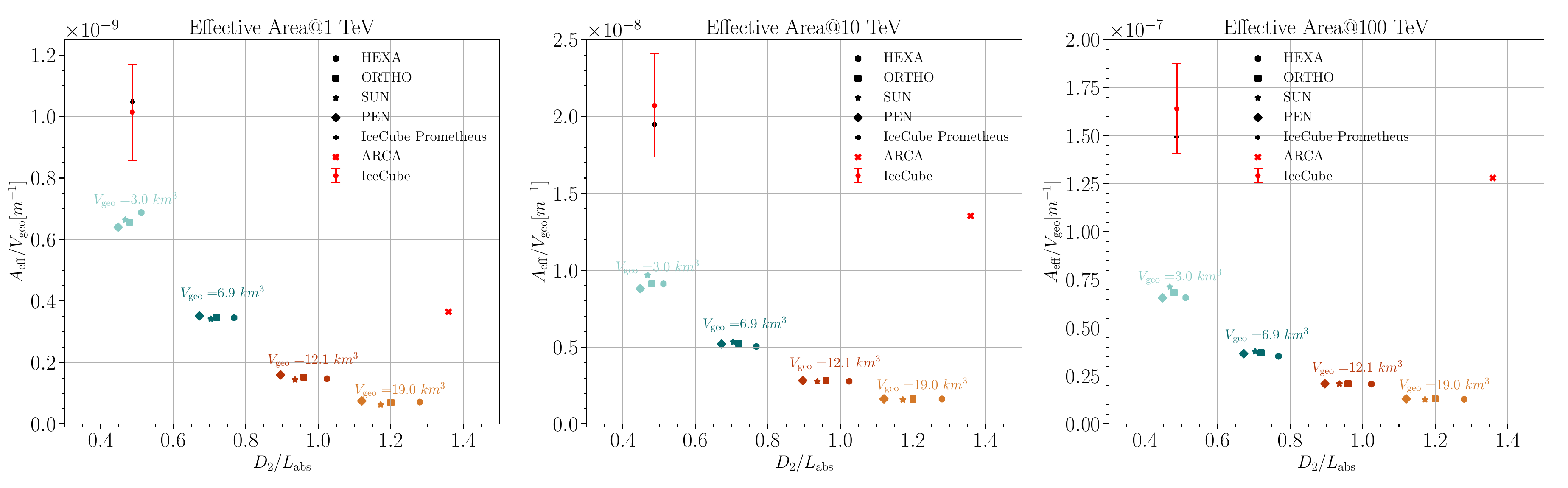}
  \caption{\textbf{Per geometrical volume effective areas of neutrino telescopes at trigger level.} Each panel shows the effective area per geometrical volume of IceCube\cite{icecubeeffa}, ARCA\cite{ARCAeffa}, IceCube simulated with \texttt{Prometheus}\cite{jeff} and our candidates at different energy, as a function of $\mathcal{D}_2$/$L_{abs}$. $\mathcal{D}_2$ is defined in~\cref{sec:datasim}, reflecting the sparsity of detector layout. $L_{abs}$ is the absorption length of the ice, introduced here to normalize neutrino telescopes of different mediums to a common reference map. Optical properties are complex and vary between experiments. To simplify, we used $L_{abs} = 250\ m$ for ice and $L_{abs} = 60\ m$ for water. Note that the trigger-level selection criteria might be different across experiments. }
  %\note{Each panel here is for a candidate shape. Colors stand for different levels of sparsity.}
  \label{fig:effa_exp}
\end{figure*}

The graphs first undergo a normalization layer to linearly scale all weights and features in a range of -1 to 1, aligning with the behavior of the activation function.
The embedding layer here applies a fully-connected layer to the node features, mapping them to a higher dimensional vector space.
The weigh-update layers implement an attention mechanism within our network.
This mechanism enables dynamic updates of attention on edges based on current attention and the features of nodes at both ends.
The graphs are then processed through four convolutional units.
Each unit contains a pooling sector that extracts information into a matrix form, which is directly connected to the final dense layers, producing the final output. 

Employing pooling after each convolutional operation helps the network retain primitive information and avoid excessive smoothing due to local blurring in convolutional layers.
In the context of angular reconstruction, the final output is a 3D vector representing the incident direction of muon neutrinos. 
If the goal is to reconstruct the initial energy of neutrino events, the output is a scalar value.

This graph neural network reconstruction algorithm, working as a generalized alternative to convolutional approaches, reaches similar performance in angular resolution: in~\Cref{subsec:benchmark}, we will show a benchmark test of the GNN algorithm.

\section{Results}\label{sec:results}

In this section, we will present the effective areas of the detector candidates at trigger and quality levels, as well as discuss the performance of both the online-capable \texttt{LineFit} algorithm on trigger-level events and the more sophisticated GNN method on quality-level events for the proposed geometries.

\subsection{Effective Areas}

We determine the effective area of each one of the sixteen geometries by using the  \texttt{LeptonWeighter} package~\cite{LI} and our simulated events, selected and computed at the two criteria levels, respectively.
The effective areas are integrated in direction, yielding an omnidirectional effective area for each of the levels of selection. In the left panel of \Cref{fig:Aeff}, we show the effective area of \texttt{hexa128} as a benchmark.
\Cref{fig:Aeff} shows the effective areas of each of the considered geometries compared to that of \texttt{hexa128}.
Detectors with similar inter-string spacing exhibit similar effective areas and energy-dependent trends.
At trigger level, $\mathcal{D}_2 \approx \SI{180}\m$ is the optimal for the in-ice candidates.
Quality event criteria filter most events towards the low-energy end of the spectrum, cutting more events there as the inter-string distance increases. However, towards the high-energy end of the flux, where we expect most of our signal to be of astrophysical instead of atmospheric origins, $\mathcal{D}_2 \approx \SI{240}\m$ yields the largest effective area overall across all detector geometries.

\begin{figure*}[t]
  \centering
  \includegraphics[width=0.75\textwidth]{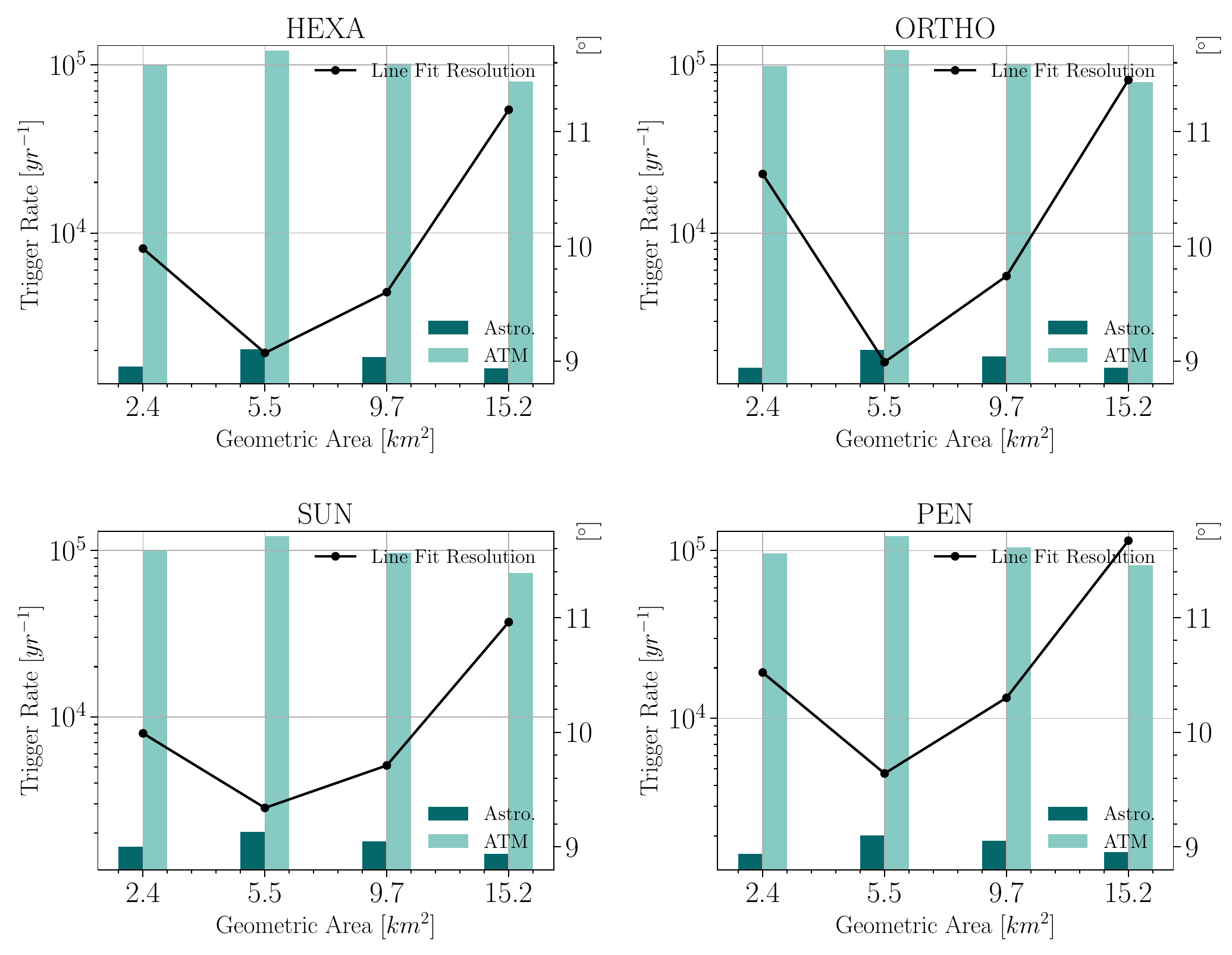}
  \caption{\textbf{Trigger rates and trigger level line-fit regression angular resolution on the simulated geometries.} Dark and light blue bars show trigger rates of astrophysical and atmospheric neutrinos correspondingly. We used astrophysical neutrino flux data from IceCube result \cite{astroresult} and atmospheric neutrino flux data from \texttt{Nuflux} \cite{nuflux}.  Black lines denote zenith resolutions of line fit reconstruction. }
  %\note{Each panel here is for a candidate shape. Colors stand for different levels of sparsity.}
  \label{fig:flux}
\end{figure*}

In~\cref{fig:effa_exp}, we show the geometrical-volume-normalized effective areas of our detector candidates in comparison with the in-ice neutrino detector IceCube~\cite{icecubeeffa} and the in-water neutrino detector ARCA~\cite{ARCAeffa}. The horizontal axis is the inter-string distances in our metric $\mathcal{D}_2$ normalized by the absorption distance in the corresponding medium; the vertical axis is the effective area normalized by geometrical volume. We have included an analysis of IceCube's effective area using simulations in \texttt{Prometheus}, and the results match the IceCube-reported results. 

From the plots, we observe that $A_\mathrm{eff}/V_\mathrm{geo}$ is lower for all our simulated candidates compared with IceCube and ARCA. This is due to our candidates expanding only sideways and thereby demonstrating more pancake-shaped geometries as opposed to the more cubic-shaped IceCube and ARCA, as previously discussed in~\Cref{sec:datasim}. For example, while our candidates with $\mathcal{D}_2 \approx 120$ meters (shown in~\Cref{fig:effa_exp} as the light green cluster of points) have the same string separations with IceCube and are all 1 km in depth, they span a geometrical area of 2.4 km$^2$ as compared to the 1 km$^2$ geometrical area of IceCube. This results in a reduced contribution from neutrinos that arrive sideways, implying that the increment in the effective area does not scale as efficiently as when both the horizontal and vertical cross-sectional geometrical area scale up equally. However, as discussed earlier, the vertical depth is largely restricted by natural conditions; therefore, a cubic design is not always feasible when aiming for an expansion in effective area.

\subsection{Trigger Level \texttt{LineFit} Resolution}

In ~\Cref{fig:flux}, we show the performance of the geometry candidates at the trigger level. For each geometry at each geometrical area, we show the expected number of atmospheric and astrophysical muon neutrino events and the mean angular resolution of the \texttt{LineFit} algorithm. For the atmospheric muon neutrino event estimation, we assume the \texttt{Honda2006} flux implemented in the \texttt{nuflux} package~\cite{nuflux}; for the astrophysical muon neutrino event number estimate, we assume a flux with normalization and spectral index taken from the IceCube cascade fit with 6 years of data in~\cite{astroresult}.
On the one hand, we see that trigger rates are highest when the geometric area of the detector reaches $A_{\textrm{geo}} =  5.5\  \si\km^2$ across all geometrical layouts, corresponding to an inter-string distance of about 180 meters, corresponding to about 0.75 times the mean absorption length of photons in ice which is approximately $d_{\mathrm{abs}} \approx 240$ meters, and the \texttt{LineFit} algorithm also performs best at this separation distance. On the other hand, there is no apparent advantage shown in any particular layout.

\begin{figure*}[t]
  \centering
  \includegraphics[width=1\textwidth]{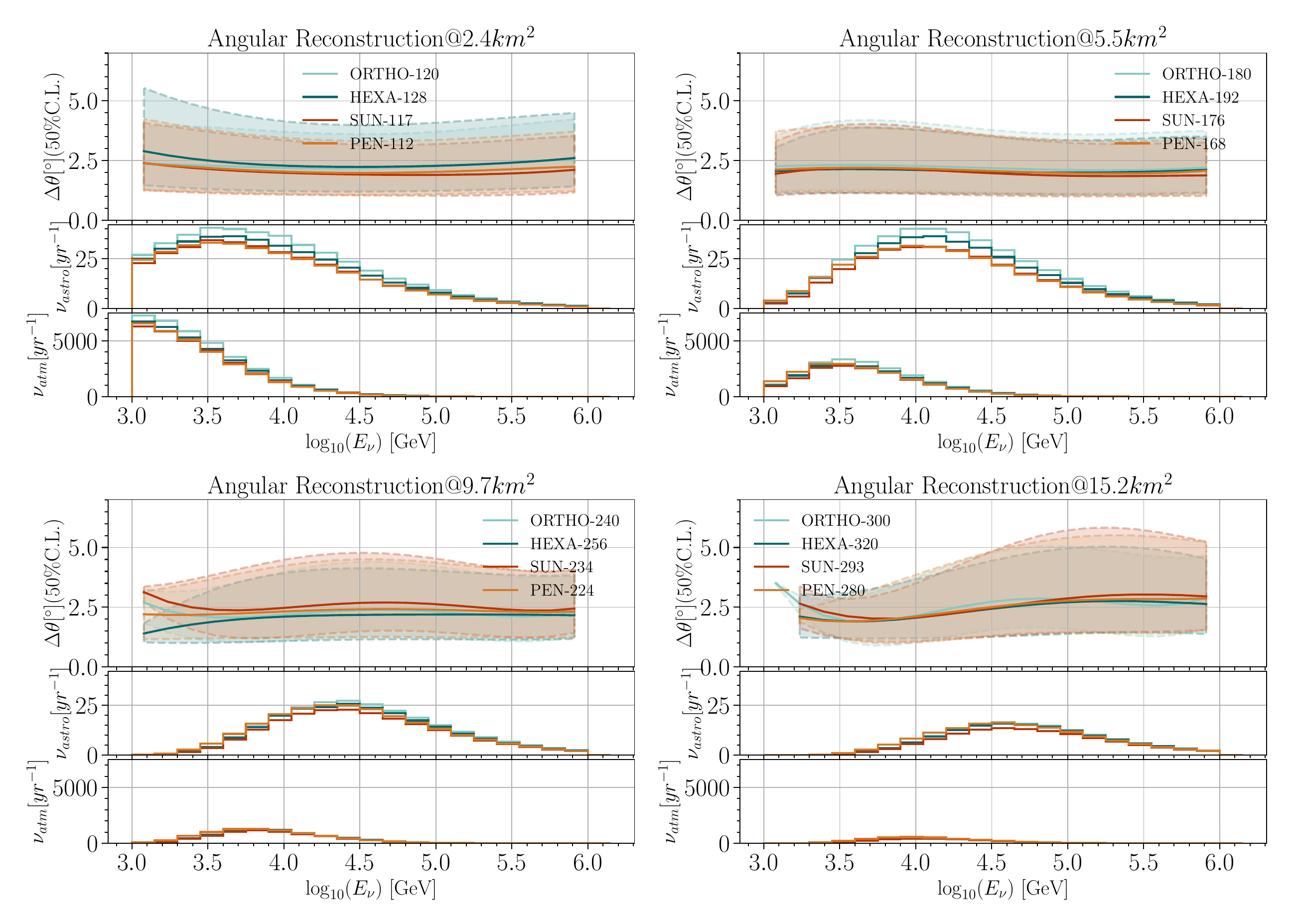}
  \caption{\textbf{Angular reconstruction performances on prototypes.} Each panel compares the different geometries at the same detector geometrical area. Each top sub-panel shows the reconstruction resolution, with the top and bottom 20\% range shown as the shaded region. The middle and bottom sub-panels show the quality-level astrophysical and atmospheric neutrino rates as a function of true neutrino energy. }
  %\note{Each panel here is for a candidate shape. Colors stand for different levels of sparsity.}
  \label{fig:angres}
\end{figure*}

\subsection{Quality-level GNN Angular Reconstruction Resolution}
At the quality level, where selected data will be analyzed with the GNN, we show the energy-binned expected number of atmospheric and astrophysical muon neutrino events in the bottom panels of~\Cref{fig:angres} as well as the angular resolution in the top panel.
As inter-string distance increases and detector geometrical volume increases, more high-energy events pass the quality selection criteria, but on the other end of the spectrum, low-energy events are not selected if the inter-string distance is too large, e.g., when stepping into the $A_{\mathrm{geo}} \geq \SI{10}\km^2$ region. This selects more astrophysical events and eliminates atmospheric background. Across all string separation distances, $10^5$ GeV is approximately the energy threshold where we start selecting more astrophysical neutrino candidates as the atmospheric background diminishes, and an inter-string distance of around $180$ to $240$ meters yields the best rate at and above this energy threshold while maintaining a decent reconstruction resolution.

We do not see a significant preference for any particular geometrical layout pattern, but we should note that the expected number of events for smaller-sized detector candidates seems to prefer a more regular layout in terms of expected event rates. However, at the trigger level, we do not see such a trend; see~\Cref{fig:Aeff}. This suggests our quality cut selection criteria, which depend on the regularity and overall shape of the event signature morphology, which prefers regular orthogonal and hexagonal layouts. Therefore, the specific geometrical layout pattern of strings should be studied in a combined manner with selection criteria as well as reconstruction methodology, whereas a simple generalized search we performed yields similar preferences across all candidates.

\section{Summary and Outlook\label{sec:summary}}
In this article, we have studied sixteen geometries grouped in four different configurations and, for each, we have considered four levels of sparsity that span from current to next-generation neutrino telescope sizes.
We are able to make some observations and conclusions using our simulations. To start with, for effective areas and their scaling, we found cubic designs with similar vertical and horizontal cross-sectional geometric areas are more "economical" than pancake-like geometries: the effective area per unit geometrical volume is larger. However, for next-generation detectors that aim at very large geometric areas, such an optimization might become impossible due to the inability to extend vertical reach. For our simulated geometries, which are all pancake-shaped, we moved on to investigate the effects of varying geometrical layouts and inter-string distances. 

In the case of inter-string distance, we found that $\mathcal{D}_2 \approx 180$ in ice medium yields the best event rate as well as angular reconstruction resolution across all geometrical layouts overall, yet for astrophysical neutrino signals, we would be looking at $\mathcal{D}_2 \approx 240$ for an optimized expected event rate: these correspond roughly to $0.72 \times$ and $0.96 \times d_{\mathrm{absorption}}$ in ice medium, respectively, aligning well with our expectation that larger inter-string distance tunes for higher energy flux.
In the case of geometrical layouts, our study found no significant overall preference for any particular string arrangement. Moreover, our results indicate that the layout should be optimized in combination with signal selection criteria, as both exert influence on signal rate; this is not to mention the necessity of studying and optimizing more specified reconstruction algorithms for the candidates, respectively. 
Furthermore, we should point out that there is no reason to restrict the geometries of next-generation telescopes in these sixteen proposed geometries.
For example, a qualitatively different study could explore uneven string spacing or non-homogeneous vertical spacing, which is motivated by rejecting incoming backgrounds.

As noted in the introduction, our results suggest that the future is not in a one-detector-to-rule-them-all ecosystem, and detectors should be optimized individually for their own discovery purposes.  
Towards this end, our work serves as a reference for future designs.
We present a generalized method to test any possible proposals using traditional (\texttt{LineFit}) and machine learning (GNN) techniques.
Combining the results of effective area and reconstruction performance, one can measure the usability of a proposed neutrino telescope for generic neutrino selection.

As a final note, while we can utilize the method in this study to find an optimal geometry, the physics reason why a specific configuration performs better remains nuanced.
This is only the first step of a complete neutrino telescope optimization, as subsequent steps require specific scientific targets.
Interpretation of a well-trained GNN might reveal the key factors that influence the reconstruction performance of a neutrino telescope, offering a new way to understand neutrino telescopes further and better design next-generation detectors.

\begin{acknowledgments}
We thank Jeffrey Lazar and Felix Yu for useful discussions.
TZ was supported by Fellowship from University of Science and Technology of China and Heising-Simons Physics Fellowship for this work. 
MJ is supported by the Faculty of Arts and Sciences of Harvard University and the National Science Foundation.
CAA are supported by the Faculty of Arts and Sciences of Harvard University, the National Science Foundation, the Research Corporation for Science Advancement, and the David \& Lucile Packard Foundation.
\end{acknowledgments}

\nocite{*}
\newpage
\bibliography{gnngeo}% Produces the bibliography via BibTeX.
\clearpage
\newpage

\onecolumngrid
\appendix

\ifx \standalonesupplemental\undefined
\setcounter{page}{1}
\setcounter{figure}{0}
\setcounter{table}{0}
\setcounter{equation}{0}
\fi

\renewcommand{\thepage}{Supplemental Methods and Tables -- S\arabic{page}}
\renewcommand{\figurename}{SUPPL. FIG.}
\renewcommand{\tablename}{SUPPL. TABLE}

\renewcommand{\theequation}{A\arabic{equation}}

\section{Graph Neural Network Details}
\input{suppl}

\end{document}

%% file: Packages.tex
\usepackage{CJKutf8}
\usepackage{diagbox} % for diagonal lines in table cells
\usepackage{tikz}
\usepackage{adjustbox}
\usepackage{amsmath}
\usepackage{graphicx}
\usepackage{fullpage}
\usepackage[colorlinks=true,allcolors=blue]{hyperref}
\usepackage[capitalise]{cleveref}
\usepackage[T1]{fontenc}
\usepackage[utf8]{inputenc}
\usepackage{siunitx}
\usepackage{array}
\input{units}

\usepackage{pifont}
\usepackage{enumitem}
\usepackage{url}
\usepackage{rotating}
\usepackage{epsfig,graphics,rotate}
\usepackage{flushend}
\usepackage{bm}
\usepackage{amssymb}
\usepackage{amsmath}
\usepackage{amsfonts}
\usepackage{gensymb}
\usepackage{booktabs}

\usepackage{microtype}
\usepackage{multirow, makecell}
\usepackage{lipsum}

\usepackage{xspace}
\usepackage{comment}
\usepackage{floatrow}
\usepackage{ragged2e}

\usepackage{nicefrac, xfrac}
\usepackage{mathtools} % for 'pmatrix' environment
\usepackage{relsize}   % for '\larger' macro

\usepackage{float}

%% file: units.tex
\DeclareSIUnit \s {\second}
\DeclareSIUnit \ns {\nano\second}
\DeclareSIUnit \mus {\micro\second}
\DeclareSIUnit \ms {\milli\second}
\DeclareSIUnit \MB {\mega\byte}
\DeclareSIUnit \GB {\giga\byte}
\DeclareSIUnit \TB {\tera\byte}
\DeclareSIUnit \PB {\peta\byte}
\DeclareSIUnit \Mbps {\mega\bit/\s}
\DeclareSIUnit \Gbps {\giga\bit/\s}
\DeclareSIUnit \Tbps {\tera\bit/\s}
\DeclareSIUnit \Pbps {\peta\bit/\s}
\DeclareSIUnit \kton {\kilo\tonne} % changed  back to kton
\DeclareSIUnit \kt {\kilo\tonne}
\DeclareSIUnit \Mt {\mega\tonne}
\DeclareSIUnit \eV {\electronvolt}
\DeclareSIUnit \keV {\kilo\electronvolt}
\DeclareSIUnit \MeV {\mega\electronvolt}
\DeclareSIUnit \GeV {\giga\electronvolt}
\DeclareSIUnit \TeV {\tera\electronvolt}
\DeclareSIUnit \PeV {\peta\electronvolt}
\DeclareSIUnit \EeV {\exa\electronvolt}
\DeclareSIUnit \m {\meter}
\DeclareSIUnit \cm {\centi\meter}
\DeclareSIUnit \in {\inchcommand}
\DeclareSIUnit \km {\kilo\meter}
\DeclareSIUnit \kV {\kilo\volt}
\DeclareSIUnit \kW {\kilo\watt}
\DeclareSIUnit \MW {\mega\watt}
\DeclareSIUnit \MHz {\mega\hertz}
\DeclareSIUnit \mrad {\milli\radian}
\DeclareSIUnit \year {years}
\DeclareSIUnit \POT {POT}
\DeclareSIUnit \sig {$\sigma$}
\DeclareSIUnit\parsec{pc}
\DeclareSIUnit\lightyear{ly}
\DeclareSIUnit\foot{ft}
\DeclareSIUnit\ft{ft}
\DeclareSIUnit \ppb{ppb}
\DeclareSIUnit \ppt{ppt}
\DeclareSIUnit \samples{S}
\DeclareSIUnit \pe{PE}

\newcommand{\enu}{\E_\enu}

%% file: Commands.tex
\definecolor{lime}{HTML}{A6CE39}
\DeclareRobustCommand{\orcidicon}{
	\begin{tikzpicture}
	\draw[lime, fill=lime] (0,0) 
	circle [radius=0.16] 
	node[white] {{\fontfamily{qag}\selectfont \tiny ID}};
	\draw[white, fill=white] (-0.0665,0.095) 
	circle [radius=0.005];
	\end{tikzpicture}
	\hspace{-2mm}
}

\foreach \x in {A, ..., Z}{\expandafter\xdef\csname orcid\x\endcsname{\noexpand\href{https://orcid.org/\csname orcidauthor\x\endcsname}
			{\noexpand\orcidicon}}
   }

%% file: suppl.tex
\subsection{Data Preprocessing and Graph Schema}\label{subsec:schema}
A crucial step in our method is the data preprocessing pipeline, which involves transforming an event into a heterogeneous graph comprising a single type of node and three types of edges.  The quality of this encoding process greatly influences how information is presented to the downstream network and ultimately affects the performance of our algorithm.

Initially, each simulation event consists of a set of collected photons with spatial and temporal information. To retain all possible information, we treat each collected photon as a node in the graph. Neutrino telescope events can naturally be interpreted as graphs in 4D-Euclidean space, with multiple nodes possessing 4D features but lacking edges.  The sizes of graphs vary from $\sim O(10)$ to $\sim O(10^5)$.

To facilitate message-passing across the graph, an essential task is to establish a scheme for generating weighted edges between nodes. While using a learnable encoder, as employed in the DYNEDGE algorithm for low-energy event reconstruction in IceCube\cite{GNN_ICLE}, initially seems promising, it becomes computationally infeasible for large graphs.  Therefore, we need to design a preprocessing approach based on our knowledge of neutrino events, connecting highly relevant photons and distant pairs in physical space that reflect global structures.

\begin{figure*}[h]
  \centering
  \includegraphics[width=0.9\textwidth]{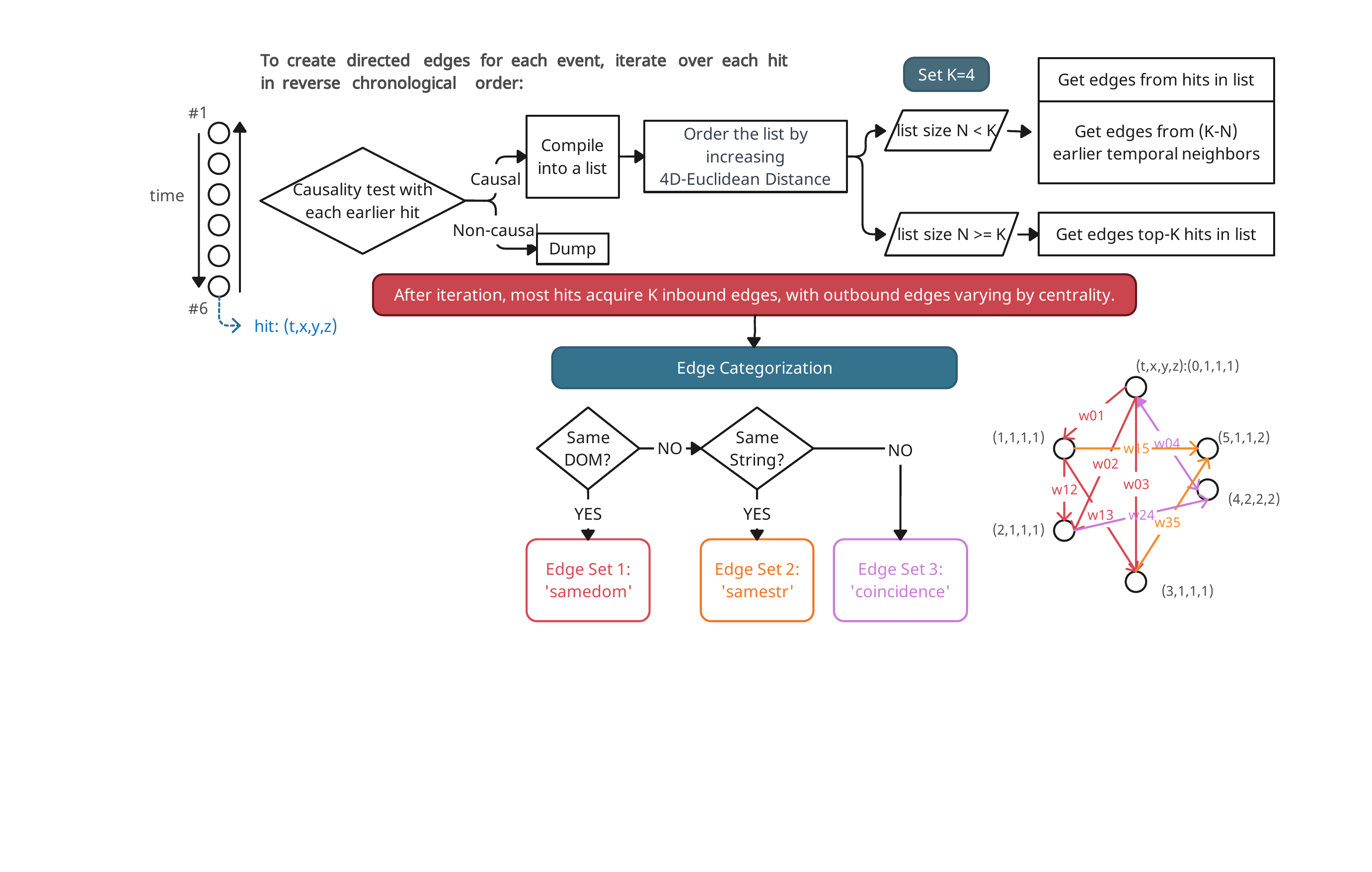}
  \caption{\textbf{Schematic diagram of preprocessing pipeline.} Side illustration shows messgae passing within the graph in each convolution step.}
  %\note{}
  \label{fig:schema}
\end{figure*}

The preprocessing algorithm begins by sorting all photons in temporal sequence, from earlier to later.  Since time flows in one direction, we establish directional edges between selected pairs, always pointing from earlier hits to later ones. This ensures that, during subsequent convolution operations, information flows along the time direction. Updated features of nodes depend only on themselves and their source neighbors. A visualization of this pipeline is shown in~\Cref{fig:schema}.

\subsection{Benchmarking Network Performance}\label{subsec:benchmark}

In this section, we show, as a benchmark test, the performance of the GNN algorithm on IceCube events and compare it against a state-of-the-art resolution algorithm to ensure its validity in performing the geometry evaluations shown previously in this work.

\begin{figure}[h]
  \centering
  \includegraphics[width=0.6\textwidth]{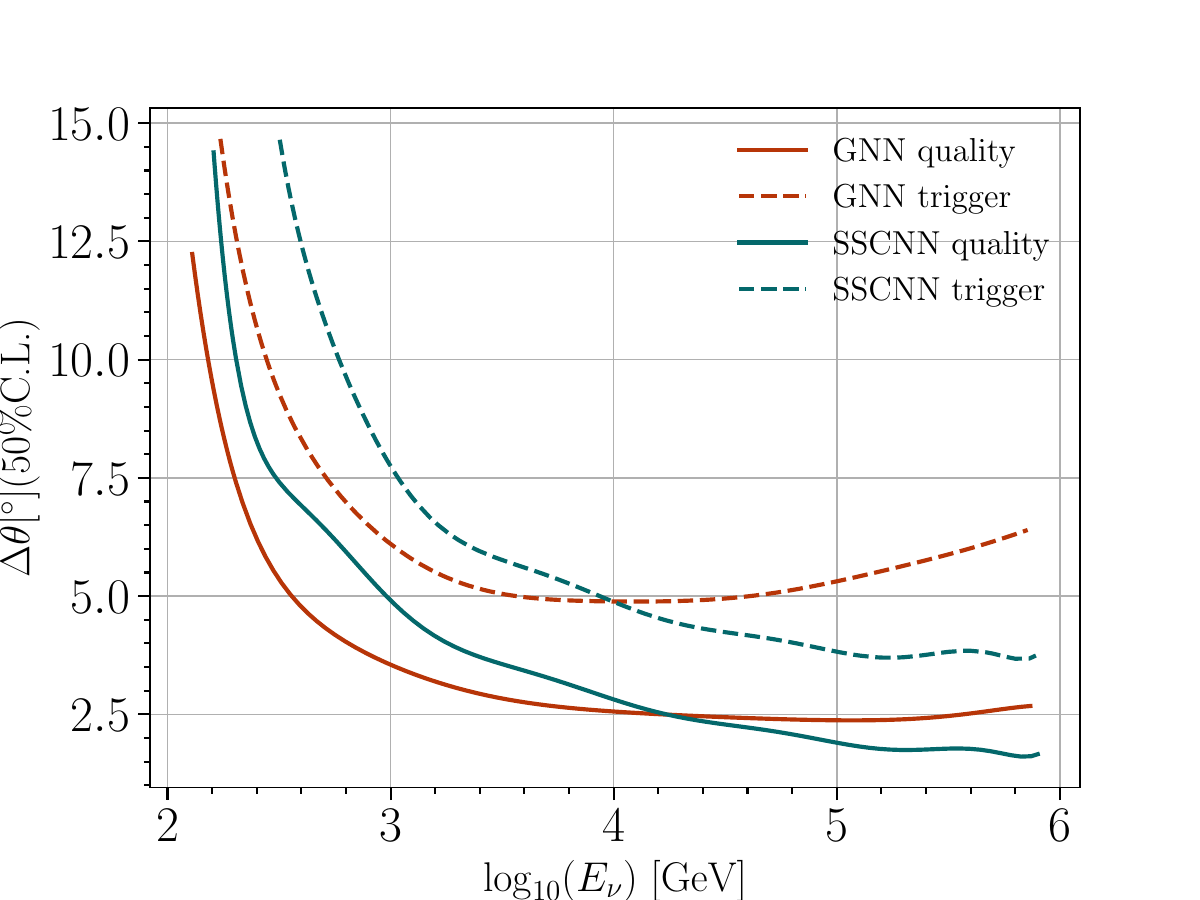}
  \caption{Angular reconstruction performance on IceCube geometry.}
  %\note{}
  \label{fig:gnnbench}
\end{figure}

In~\Cref{fig:gnnbench}, the performance of our proposed method is showcased through a comparison with the SSCNN method\cite{felix}, revealing compelling results in the context of IceCube simulation events. Here and in the following results, we use a classic metric for the relevant tasks, the median angular difference, to measure the performance of angular reconstruction, defined as the 50\% quantile of angular differences between true and predicted vectors.

The GNN algorithm demonstrates good performance in the low-energy range, outperforming the SSCNN method. However, as the energy surpasses 10 TeV, our method exhibits a slightly diminished performance. This observation leads us to speculate that the presence of corner clippers, events that are not fully contained within the detector, may contribute to the upgoing trend at the high-energy end. By applying quality cuts, we can observe a smoothing effect on the upward trend.